\documentclass[10pt]{article}
\usepackage{natbib}
\usepackage{amssymb,amsmath}
\usepackage{graphicx}
\usepackage{color}
\usepackage{hyperref}
\hypersetup{
	colorlinks=true,
	linkcolor=black,
	filecolor=red,      
	citecolor=blue,
}
\usepackage[dvipsnames]{xcolor}
\definecolor{myblue}{RGB}{31,119,180}
\definecolor{myorange}{RGB}{255,127,14}
\definecolor{mymagenta}{RGB}{255,0,255}
\definecolor{mygreen}{RGB}{0,128,0}
\usepackage{curves,multicol}
\usepackage{geometry}
\usepackage{times}
\usepackage{bm}
\usepackage{ulem}
\usepackage{ucs}
\usepackage[utf8x]{inputenc}
\usepackage[intoc]{nomencl}
\usepackage{float}
\usepackage{mathrsfs}
\usepackage{cleveref}
\usepackage{units}
\usepackage{titlesec}
\usepackage[labelfont=bf]{caption}
\usepackage{lipsum}   
\usepackage[none]{hyphenat} 
\usepackage{makecell}
\usepackage{tikz}

\usepackage{scalerel,stackengine}

\stackMath
\newcommand\reallywidehat[1]{%
	\savestack{\tmpbox}{\stretchto{%
			\scaleto{%
				\scalerel*[\widthof{\ensuremath{#1}}]{\kern-.6pt\bigwedge\kern-.6pt}%
				{\rule[-\textheight/2]{1ex}{\textheight}}
			}{\textheight}%
		}{0.5ex}}%
	\stackon[1pt]{#1}{\tmpbox}%
}
\crefname{equation}{Equation}{Equations}
\Crefname{equation}{Equation}{Equations}
\Crefname{figure}{Figure}{Figures}
\crefname{table}{Table}{Tables}
\Crefname{tabular}{Table}{Tables}

\def\dashdotted{\xleaders\hbox to 1em{$- \cdot$}\hfill $-$}

\newcommand*\sq{\mathbin{\vcenter{\hbox{\rule{.27ex}{.27ex}}}}}

\makeatletter

\renewcommand\@biblabel[1]{}

\renewcommand{\section}{\@startsection
{section}
{1}
{0mm}
{-\baselineskip}
{0.5\baselineskip}
{\normalfont\bfseries\MakeUppercase}} 

\renewcommand{\subsection}{\@startsection
{subsection}
{2}
{0mm}
{0.5\baselineskip}
{0.25\baselineskip}
{\bfseries\normalsize}} 

\makeatother

\begin{document}
\sloppy
\pagenumbering{arabic}
\setcounter{secnumdepth}{-1} 

\vspace*{-1.0cm}
\begin{flushright} \vbox{
33rd Symposium on Naval Hydrodynamics\\
Osaka, Japan, 18-23 October 2020}
\end{flushright}

\vskip0.65cm
\begin{center}
\textbf{\LARGE
Study of Nonlinear Interaction between Waves and Ocean Currents Using High-Fidelity Simulation and Machine Learning\\[0.35cm]
}

\Large Tianyi Li, Anqing Xuan, Lian Shen\\ 

(St. Anthony Falls Laboratory and Department of Mechanical Engineering, University of Minnesota, Minneapolis, MN 55414, USA)\\
\vspace*{0.25cm}

\end{center}

\begin{multicols*}{2}

\section{Abstract}

Modeling ocean surface waves under complex ocean current conditions is of crucial importance to many naval applications. For example, traveling ships and underwater vehicles generate spatially heterogeneous currents behind them through their drag and propeller motions. The strong currents can influence the surface wave pattern in the ship wake. In this study, the nonlinear interactions between waves and complex wake currents are investigated using numerical simulations. An in-house code is developed for high-fidelity simulations of a nonlinear phase-resolved ocean wavefield interacting with subsurface currents. Several typical wake patterns are simulated using the present numerical method, and the influence of complex currents on the waves is analyzed quantitatively using theoretical solutions of wave--current interactions. We also present a method for solving the inverse problem of deducing the current field based on surface-wave data using machine-learning techniques. A deep neural network is designed for processing spatial-temporal surface wave data. Detailed analyses on the distributions of regression errors and the training dataset-dependency show that the proposed neural network can effectively deduce the current field.

\section{Introduction}
\label{SECintroduction}
In oceans, complex current motions can have significant impacts on wave dynamics. Currents can be generated by natural causes, such as the rotation of the Earth, by wind shear, and by human activities, such as surface and underwater vehicles. Directly capturing wave dynamics under complex current conditions is beneficial to revealing the mechanisms of wave--current interactions and can be applied to many engineering applications. However, the physics of irregular broadband ocean waves propagating on currents are complex. Wave--current interactions modulate wave heights; wave refraction by currents modifies the wave propagating direction; and the Doppler effect alters the dispersion relationship of surface waves. Moreover, currents are usually spatially heterogeneous. 

With the rapid development of computational capacities, numerical simulations have become an increasingly important tool in the studies of ocean waves under realistic ocean environmental conditions. A variety of algorithms have been developed to simulate ocean waves, including the higher-order spectral method~\citep{dommermuth1987high,west1987new}, the single-phase free-surface flow solver based on the Navier--Stokes equations~\citep{xuan2019conservative}, multi-phase flow solvers~\citep{weymouth2010conservative,fu2013detailed,yang2018direct}, and a boundary integral based wave--current solver~\citep{nwogu2009interaction,wang2018fully}. In this study, we use our in-house code, which utilizes the velocity-based boundary integral algorithm~\citep{nwogu2009interaction}, to conduct numerical investigations of the interactions between irregular broadband waves and complex currents in the ship wake.

Considering that the simulation of wave dynamics in ocean current conditions is a forward problem, we are also interested in the inverse problem, namely the detection of current distributions from surface-wave data. Near-surface ocean currents are difficult to measure accurately due to the complex marine environments in the field, while the surface waves are relatively easy to characterize using remote sensing techniques, such as high-frequency radar. Traditionally, Doppler effects extracted from the surface wave spectrum are utilized to model current distributions underneath. The algorithm to deduce an ocean current based on the extraction of the Doppler effects was first developed by \citet{crombie1955doppler}. Different algorithms to construct current velocity as a function of depth have been proposed, such as the effective depth method~\citep{fernandez1996measurements}, the Laplacian transform-based inversion method~\citep{ha1979remote}, and the recent polynomial effective depth method~\citep{smeltzer2019improved}. All these methods aim to deduce depth-varying current velocity from a measured phase velocity of the surface waves in a wide range of wave numbers based on the theoretical dispersion relation of waves above currents. However, to accurately deduce a current profile using these methods is still challenging because strictly speaking, the inverse problem is ill-posed mathematically. Alternatively, data-driven methods, such as deep neural networks, show the potential to handle inverse problems~\citep{adler2017solving}. In the data-driven approach, \textit{a priori} knowledge of the dispersion properties of waves is not needed. Given a sufficiently large dataset containing the information of surface waves and current distributions, one can train a deep neural network to generate an optimal model by minimizing the designed loss function. The advantages of the data-driven model include easy generalization and low computational cost for operations. The proposed neural network in this study can be easily generalized to complex ocean environments, such as spatially heterogeneous currents. In addition, once trained, the model is computationally efficient enough to operate for real-time measurements.

The remainder of this paper is organized as follows. First, the forward problem to numerically investigate ocean wave dynamics under complex ocean current conditions is discussed. Then, the design of the machine-learning algorithm for the inverse problem of deducing current distribution from surface waves is introduced. Last, conclusions are given.

\section{Phase-resolved Wave--current \\ Simulation}
In this section, we first introduce the numerical method of the simulation code developed in house for the high-fidelity simulation of the three-dimensional wavefield with the wave phases resolved and the nonlinear wave-wave interaction and wave--current interaction captured. This simulation tool can compute the broadband waves interacting with arbitrary spatially heterogeneous ocean currents. Then, we use the code to investigate the modulation effects of horizontally sheared currents and vortical flows on broadband irregular surface waves.

\subsection{Governing Equations}
We denote the Cartesian coordinates as $(x,y,z)$, where $x$ and $y$ are horizontal coordinates, and $+z$-axis points upward with the plane $z=0$ located at the mean ocean surface level. Let $(U,V,W)$ denote the ocean current velocity, $(u,v,w)$ be the wave-induced velocity, and $\eta$ be the surface elevation. The current velocity can be assumed to be steady, considering that the timescale of the current evolution is much longer than the timescale of wave motions. The surface elevation $\eta$ and scaled surface tangential velocity components $(u_s, v_s)$ defined as 
\begin{align}
u_s=u|_{z=\eta}+w|_{z=\eta}\eta_x\\
v_s=v|_{z=\eta}+w|_{z=\eta}\eta_y
\end{align}
are chosen to be the primary variables. Derived from the Euler's equations, the governing equations for $\eta$, $u_s$, and $v_s$ are~\citep{nwogu2009interaction}
\begin{align}
&\eta_t=u_n-(U_\eta\eta)_x-(V_\eta\eta)_y\label{eq:gov1}
\end{align}
\begin{align}
&u_{s,t}+\left[g\eta+\frac{1}{2}\left(u_s^2+v_s^2\right)-\frac{1}{2}w_\eta^2(1+\eta_x^2+\eta_y^2)\right]_x\notag\\&-(v_s-w_\eta\eta_y)(v_{s,x}-u_{s,y})+w_\eta U_\eta^\prime\notag\\
&+U_\eta u_{s,x}+V_\eta u_{s,y}=0,\label{eq:gov2}
\end{align}
\begin{align}
&v_{s,t}+\left[g\eta+\frac{1}{2}\left(u_s^2+v_s^2\right)-\frac{1}{2}w_\eta^2(1+\eta_x^2+\eta_y^2)\right]_y\notag\\
&+(u_s-w_\eta\eta_x)(v_{s,x}-u_{s,y})+w_\eta V_\eta^\prime\notag\\
&+U_\eta u_{s,x}+V_\eta u_{s,y}=0,\label{eq:gov3}
\end{align}
where $(U_\eta,\,V_\eta)=(U|_{z=\eta},\,V|_{z=\eta})$ are the current velocity components at the free surface, $(U'_\eta,\,V'_\eta)$ are their vertical derivatives at the surface, and $w_\eta=w|_{z=\eta}$ is the normal velocity at the surface. \Cref{eq:gov1,eq:gov2,eq:gov3} are not closed, and a velocity-based boundary-integral equation for the scaled normal-wave velocity $u_n=w_\eta-u_\eta\eta_x-v_\eta\eta_y$ is used to obtain a closure system~\citep{nwogu2009interaction}. The boundary-integral equation can be expanded into Fourier-based series when the wave steepness is chosen as the perturbation parameter, and the leading order terms are kept to resolve nonlinear wave--wave interactions. The governing equation for the normal wave velocity $u_n$ can be written as
\begin{align}
u_n=\mathcal L^{(1)}\{\bm u_s,\eta\}+\mathcal L^{(2)}\{u_n,\eta\}+\mathcal L^{(3)}\{\bm \Omega,\eta\}.\label{eq:def_un}
\end{align}
Here, the operators $\mathcal L^{(1)}\{\bm u_s,\eta\}$ and  $\mathcal L^{(2)}\{u_n,\eta\}$ are defined as
\begin{align}
\mathcal L^{(1)}\{\bm u_s,\eta\}=
&-\mathcal F^{-1}\left\{\left(\frac{i\bm{k}}{k}\right)\cdot\mathcal F\left\{\bm u_s\right\}\right\}\notag\\
&-\frac{1}{2}\eta^2\left\{\mathcal F^{-1}ik\bm{k}\cdot\mathcal{F}\left\{\bm u_s\right\}\right\}\notag\\
&+\eta\mathcal{F}^{-1}\left\{ik\bm{k}\cdot\mathcal{F}\left\{\eta\bm u_s\right\}\right\}\notag\\
&-\frac{1}{2}\mathcal F^{-1}\left\{ik\bm k\cdot \mathcal F\left\{\eta^2 \bm u_s\right\}\right\}\notag\\
&-\eta\nabla_h\eta\cdot\mathcal F^{-1}\left\{k\mathcal F\left\{\bm u_s\right\}\right\} \notag\\
&+\nabla_h\eta\cdot\mathcal F^{-1}\left\{k\mathcal F\left\{\eta \bm u_s\right\}\right\}\notag\\
&-\mathcal F^{-1}\left\{\left(\frac{i\bm k}{k}\right)\cdot\mathcal F\left\{\bm u_s \times \nabla_h\eta\right\}\right\},\label{eq:def_l1us}
\end{align}
\begin{align}
\mathcal L^{(2)}\{u_n,\eta\}=&-\nabla_h \eta\cdot\mathcal{F}^{-1}\left\{\left(\frac{i\bm k}{k}\right)\mathcal F\left\{u_n\right\}\right\}\notag\\
&+\eta\mathcal F^{-1}\left\{k\mathcal F\left\{u_n\right\}\right\}\notag\\
&-\mathcal F^{-1}\left\{k\mathcal F\left\{\eta u_n\right\}\right\},\label{eq:def_l2un}
\end{align}
where $k=|\bm k|$ is the magnitude of the wavenumber vector $\bm{k}$ and $\nabla_h=(\partial_x, \partial_y)$ denotes the horizontal gradient. Operators $\mathcal{F}$ and $\mathcal{F}^{-1}$ denote the Fourier transform and the inverse transform, respectively.
The operator $\mathcal L^{(3)}\left\{\bm\Omega,\eta\right\}$ is related to the current vorticity $\bm\Omega=(-V_z, U_z,0)$ and the surface elevation $\eta$, and is defined as
\begin{align}
\mathcal L^{(3)}\{\Omega,\eta\}=&2\left(\int_V\bm\Omega(\bm x^\prime)\times\nabla^\prime G(\bm x;\bm x^\prime)\mathrm{d}\bm x^\prime\right)\notag\\
&\cdot\left(\partial_x\eta,\partial_y\eta,-1\right),
\end{align}
where $\nabla^\prime G(\bm x;\bm x^\prime)$ denotes the gradient of the three-dimensional Green's function with respect to the variable $\bm{x^\prime}$, and $V$ is the water domain.
\subsection{Numerical Schemes}
\Cref{eq:gov1,eq:gov2,eq:gov3} are evolution equations for variables $\eta$, $u_s$, and $v_s$, respectively, and the fourth-order Runge--Kutta method is used for their time advancement. The Fourier-based pseudo-spectral method is adopted to compute spatial derivatives, i.e., $\partial_x$ and $\partial_y$, and we use the $3/2$ rule to eliminate the aliasing error of the nonlinear terms. The boundary integral equation for the wave normal velocity $u_n$ (\cref{eq:def_un}) is highly nonlinear, and the following iteration scheme is used to solve for $u_n$,
\begin{align}
u_n^{(m+1)}=\mathcal L^{(1)}\{\bm u_s,\eta\}+\mathcal L^{(2)}\{u_n^{(m)},\eta\}+\mathcal L^{(3)}\{\bm \Omega,\eta\}.\label{eq:itr_un}
\end{align}
Given the initial guess $u_n^{(0)}=0$, \cref{eq:itr_un} can be evaluated iteratively for $\{m=0,1,\cdots\}$ till the solution converges. In the present study, the convergence criterion is set to be when the mean square error between two consecutive iterations, $u_n^{(m+1)}$ and $u_n^{(m)}$, is smaller than the threshold $\epsilon=10^{-6}$. Note that when $\eta\rightarrow0$ and no currents are present, \cref{eq:def_un} reduces to the Airy wave solution. Based on the assumption that the wave steepness is within the perturbative regime, we can expect that the iteration scheme \cref{eq:itr_un} has a converged solution~\citep{li2020safe}. Due to the nonlinear wave--wave interaction, the wave energy at the higher frequencies would keep increasing without dissipation. Therefore, a low-pass filter is applied to $\eta$, $u_s$, $v_s$, and $u_n$ in the wavenumber space to introduce numerical dissipation and increase the numerical stability~\citep{xiao2013rogue}. The filter can be written as a Fourier multiplier $\Lambda$ acting on a function $f(x,y)$,
\begin{align}
\mathcal{F}\left\{\Lambda f\right\}(\bm k)= \exp\left[-\left(\frac{|\bm k |}{\beta_1 k_p}\right)^{\beta_2}\right]\mathcal{F}\left\{f\right\}(\bm k),
\end{align}
where $k_p=\arg\max_{|\bm k|}|\mathcal{F}(f)|$ is the peak wavenumber of the wave spectrum and the constants $\beta_1=8$ and $\beta_2=30$.

\subsection{Problem Setup}
To study how current motions modulate the dynamics of realistic surface waves, we use the wave spectrum~\citep{hasselmann1973measurements} from the Joint North Sear Wave Project (JONSWAP) to initialize the undisturbed ocean wavefield. The omni-directional frequency spectrum of the wavefield, $E(\omega)$ with $\omega$ being the wave angular frequency, is given by
\begin{align}
E(\omega)=\frac{\alpha_p g^2}{ \omega^5}\exp\left[-\frac{5}{4}\left(\frac{\omega}{\omega_p}\right)^{-4}\right]\gamma^{\exp\left[-\frac{(\omega-\omega_p)^2}{2\sigma^2\omega_p^2}\right]}.\label{eq:jonswap}
\end{align}
Here, the constant $g=9.8\,\mathrm{m/s^2}$ is the gravitational acceleration, $\alpha$ is the parameter associated with the total wave energy, $\omega_p$ is the peak angular wave frequency, $\gamma=3.3$ is a dimensionless constant, and $\sigma$ is set as~\citep{hasselmann1973measurements}
\begin{align}
\sigma=
\begin{cases}
0.07 & \omega\leq \omega_p,\\
0.09 & \omega>\omega_p.
\end{cases}
\end{align}
The following empirical formulae are used to parameterize $\alpha_p$ and $\omega_p$,
\begin{align}
\alpha=0.076\left(\frac{U_{10}^2}{Fg}\right)^{0.22},\label{eq:alpha}\\
\omega_p=22\left(\frac{g^2}{U_{10}F}\right)^{1/3},\label{eq:omega}
\end{align}
where $U_{10}$ is the wind velocity at 10 meters above the sea surface and $F$ is the fetch. For the simulations considered in this work, we set $U_{10}=6\,\mathrm{m/s}$ and $F=10\,\mathrm{km}$, resulting in $\alpha=0.0133$ and $\omega_p=2.57\,\mathrm{s}^{-1}$. To obtain the directional wave spectrum $E(\omega,\theta)$, we multiply a spreading function $D(\theta)$ to \cref{eq:jonswap}~\citep{longuet1963effect}
\begin{align}
E(\omega,\theta)=E(\omega)D(\theta),\label{eq:jonswap_2d}
\end{align}
where the spreading function is chosen as $D(\theta)=(2/\pi)\cos^2(\theta)$, $\theta\in[-\pi/2,\pi/2)$. The parameters for initializing the two-dimensional JONSWAP wave field are summarized in  \cref{TAB_JONSWAP}. 
\begin{table}[H]
	\caption{Summary of JONSWAP wave parameters}
	\begin{small}
		\begin{center}
			\begin{tabular}{|c|c|c|c|c|}
				\hline
				$\alpha$ & $\omega_p~(\mathrm{s}^{-1})$ & $\lambda_p~(\mathrm{m})$ & $c_p~(\mathrm{m/s})$ &$T_p~(\mathrm{s})$\\
				\hline
				$0.0133$& $2.57$ & $9.32$& $3.81$ & $2.45$\\
				\hline
			\end{tabular}
		\end{center}
	\end{small}
	\label{TAB_JONSWAP}
\end{table} 
The domain size for the wave field is $L_x\times L_y=200\,\mathrm{m}\times200\,\mathrm{m}$, which is large enough for capturing all the dynamical interactions between ship-induced currents and ocean waves. A grid with $N_x\times N_y=1024\times 1024$ is adopted to discretize the horizontal domain, corresponding to a grid spacing of $\Delta_x=\Delta_y=0.195\,\mathrm{m}$. Therefore, one peak wavelength $\lambda_p$ (\cref{TAB_JONSWAP}) is well resolved by 48 discrete grid points. For the time discretization, we set the time step $\Delta_t=0.049\,\mathrm{s}$, which is $1/50$ of the peak wave period.

\begin{table}[H]
	\caption{Summary of simulation parameters}
	\begin{small}
		\begin{center}
			\begin{tabular}{|c|c|c|c|c|}
				\hline
				$L_x~(\mathrm{m})$ & $L_y~(\mathrm{m})$ &$\Delta_x~(\mathrm{m})$ &$\Delta_y~(\mathrm{m})$ & $\Delta_t~(\mathrm{s})$ \\
				\hline
				$200$ & $200$ & $0.195$&$0.195$& $0.049$\\
				\hline
			\end{tabular}
		\end{center}
	\end{small}
	\label{TABexampletable}
\end{table}

\begin{figure}[H]
	\centering
	\includegraphics[width=0.95\columnwidth]{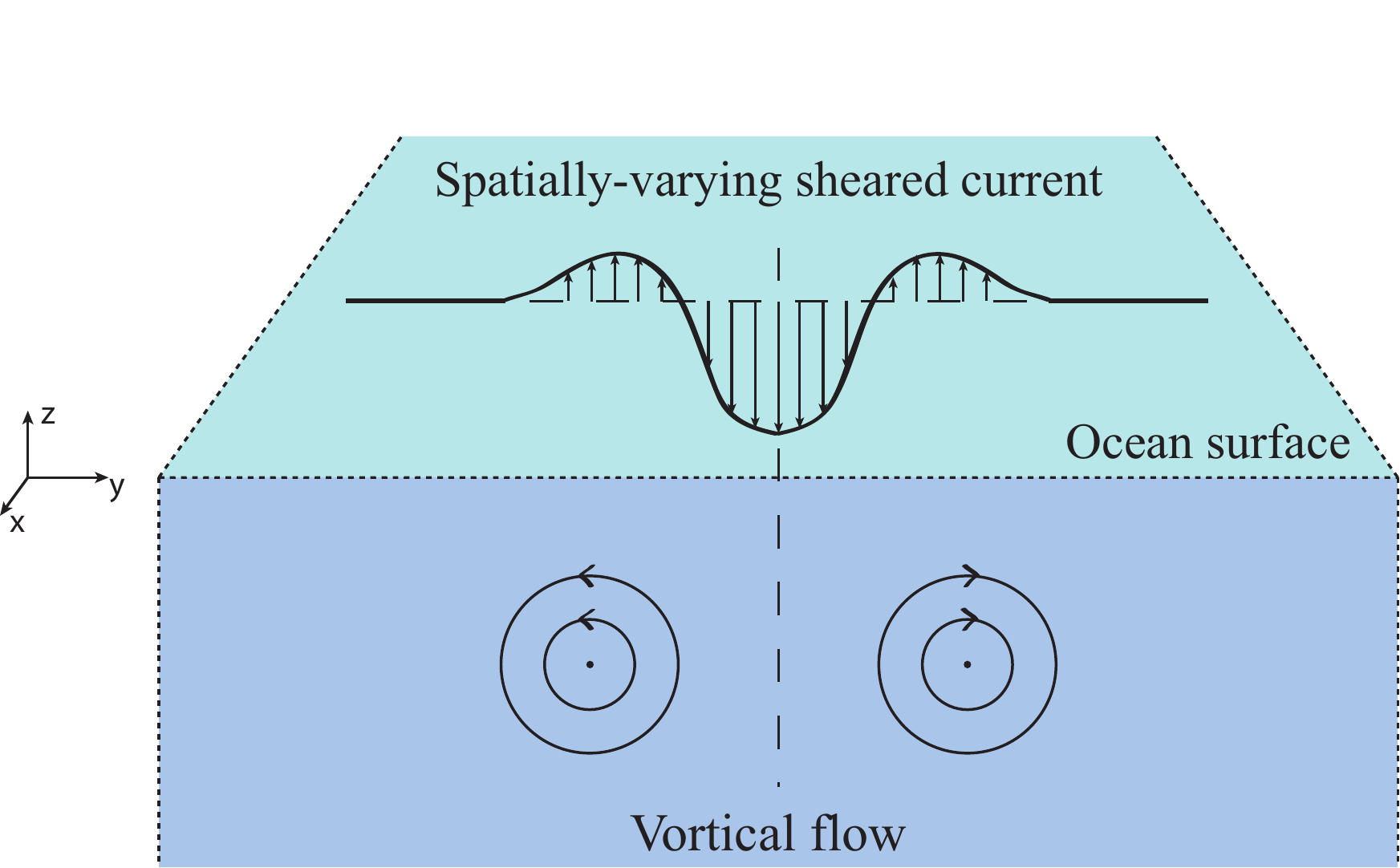}
	\caption{Configuration of ship-induced current velocity distribution in the wake.} \label{fig:current_config}
\end{figure}

There are two main mechanisms for current generation by a ship. A traveling ship drags the surface water, resulting in a horizontally-varying current in the moving direction of the ship. Meanwhile, ship propellers can generate rotating flows~\citep{somero2018structure}. \Cref{fig:current_config} shows the configuration of the current velocity field in the ship wake, which consists of the spatially varying horizontal velocity induced by the drag and the vortical motions induced by the propellers.
 
The drag-induced current velocity $U_d$ is modeled by the superposition of three Gaussian profiles as
\begin{align}
U_d(x,y,z)=\frac{U_{d0}L_0}{\sqrt{2\pi}\sigma_m}\exp\left(-\frac{1}{2}\left(\frac{y-L_m}{\sigma_m}\right)^2\right)\notag\\
-\frac{U_{d0}L_0}{\sqrt{2\pi}\sigma_l}\exp\left(-\frac{1}{2}\left(\frac{y-L_l}{\sigma_l}\right)^2\right)\notag\\
-\frac{U_{d0}L_0}{\sqrt{2\pi}\sigma_r}\exp\left(-\frac{1}{2}\left(\frac{y-L_r}{\sigma_r}\right)^2\right).\label{eq:Ud}
\end{align}
The current velocity distribution described by \Cref{eq:Ud} varies smoothly in the $y-$direction and is uniform in the $x-$ and $z-$directions. The mean-current velocity in the entire domain is zero. Setting the origin of the coordinates at one corner of the rectangular domain, we choose the constants in \cref{eq:Ud} as $L_m=100\,\mathrm{m}$, $L_l=110\,\mathrm{m}$, $L_r=90\,\mathrm{m}$, $L_0=1\,\mathrm{m}$, $\sigma_m=8\,\mathrm{m}$, and $\sigma_l=\sigma_r=6\,\mathrm{m}$. The characteristic velocity $U_{d0}$ is chosen such that the maximum drag-induced surface current velocity is $1\,\mathrm{m/s}.$

Propeller-induced rotating flows are modeled using two counter-rotating line vortices, which have the following stream function,
\begin{align}
\psi=\psi_r+\psi_l=&-\frac{\Gamma}{2\pi}\left(\ln\sqrt{(z-z_r)^2+(y-y_r)^2}\right.\notag\\
&\left.-\ln\sqrt{(z-z_l)^2+(y-y_l)^2}\right).
\label{eq:vortex_Psi}
\end{align}
The centers of the two vortices, $(y_l,z_l)$ and $(y_r,z_r)$, are located symmetrically about the centerline of the domain $y=L_m$ at the same depth. The distance between the vortex cores is set as $|y_l-y_r|=10\,\mathrm{m}$ and the depth is $z_l=z_r=-5\,\mathrm{m}$. The current velocity can be obtained by calculating $(U,V,W)=(0,\partial_z\psi,-\partial_y\psi)$. At the free surface, the propeller-induced rotating currents have the following horizontal-velocity distribution,
\begin{align}
V_{p\eta}=\frac{\Gamma}{2\pi}\left(\frac{y_l}{\sqrt{(y-y_l)^2+z_d^2}}-\frac{y_r}{\sqrt{(y+y_r)^2+z_r^2}}\right)
\end{align}
The vortex strength $\Gamma$ is chosen such that the maximum of propeller-induced velocity at the free surface is $0.1\,\mathrm{m/s}$. The sign of $\Gamma$ represents two different propeller working conditions. The condition where $\Gamma>0$ represents inward rotating propellers, which result in a converging flow near the centerline of the ship-wake region. When $\Gamma<0$, propellers rotate outward, and the surface flow diverges from the centerline. 
The horizontal-varying sheared-flow \cref{eq:Ud} and the vortical-flow~\cref{eq:vortex_Psi} together construct the wake velocity.
In this study, we set up three cases, including the case without propellers' rotating effects (NR), the case with outward rotating propellers (OR), and the case with inward rotating propellers (IR). The details of the simulation cases are summarized in \cref{tab:casename}.

\begin{table}[H]
	\caption{Summary of simulation parameters}
	\begin{small}
		\begin{center}
			\begin{tabular}{|c|c|c|c|}
				\hline
				Case & Propeller Rotation & \makecell{$\max U_\eta$ \\ $(\mathrm{m/s})$} & \makecell{$\max V_\eta$ \\ $(\mathrm{m/s})$} \\
				\hline
				NR & No  & $1.0$ & $0.0$\\
				\hline
				OR & Outwards Rotating & $1.0$ & $0.1$\\
				\hline
				IR & Inwards Rotating & $1.0$ & $0.1$\\
				\hline
			\end{tabular}
		\end{center}
	\end{small}
	\label{tab:casename}
\end{table} 

We note that because we focus on the wave dynamics in the presence of ship-wake flows, the ship-generated wake currents are kept steady in the simulation to facilitate the quantitative analyses of the current effect on waves. In addition, for the present configuration, the spatial and temporal decay of wake flows can be considered negligibly small within the duration of the simulation because the time scale of the surface waves is much smaller than the time scale of the current decay. Under the assumption that the wake decay is negligible, the current velocity is set to be uniform in the streamwise direction.

\subsection{Results}
\begin{figure}[H]
	\includegraphics[width=0.95\columnwidth]{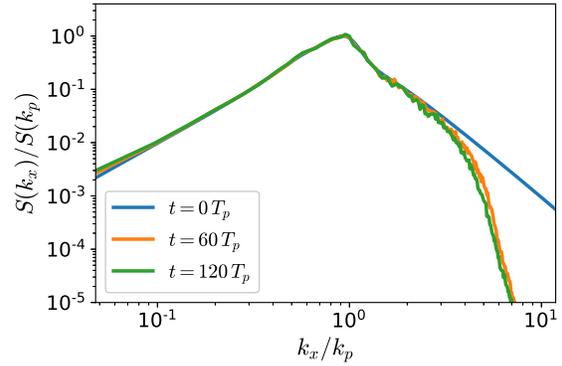}
	\caption{Evolution of the wave energy spectrum $S(k_x)$.} \label{fig:jonswap_init}
\end{figure}

The variables $\eta$, $u_s$, and $v_s$ of the initial wavefield at $t=0$ are calculated using the superposition of linear waves with different frequencies following the directional JONSWAP wave spectrum (\cref{eq:jonswap_2d}). To let the nonlinear wave dynamics develop, the simulation is first run with no current-velocity distributions for $120$ peak-wave periods. \Cref{fig:jonswap_init} shows the temporal evolution of the one-dimensional wave-energy spectrum, defined as
\begin{align}
S(k_x)=\frac{1}{N_y}\sum_{i=1}^{N_y}\left|\reallywidehat{\eta}^x(k_x,y_i)\right|^2,
\end{align}
where $\reallywidehat{\eta}^x (k_x,y)$ denotes the Fourier transform of $\eta(x,y)$ in the $x$-direction.

As shown in \Cref{fig:jonswap_init}, wave energy grows with time in the low wavenumber region, $k\ll k_p$, owing to the inverse cascade of wave energy~\citep{zakharov1992inverse}. On the other hand, in the high wavenumber region, $k\gg k_p$, the wave loses energy to dissipation~\citep{xiao2013rogue}. The instantaneous wave field at $t=120\,T_p$ is used as the initial condition of the wave--current interaction simulations. The current velocity $(U,V,W)$ is introduced at $t>120\,T_p$. The simulations are run for an additional $40\,T_p$ to obtain converged results of the wave dynamics. The wave data at the time interval $160T_p<t<200T_p$ is collected to analyze the interaction between surface waves and subsurface currents.

\begin{figure}[H]
	\includegraphics[width=0.95\columnwidth]{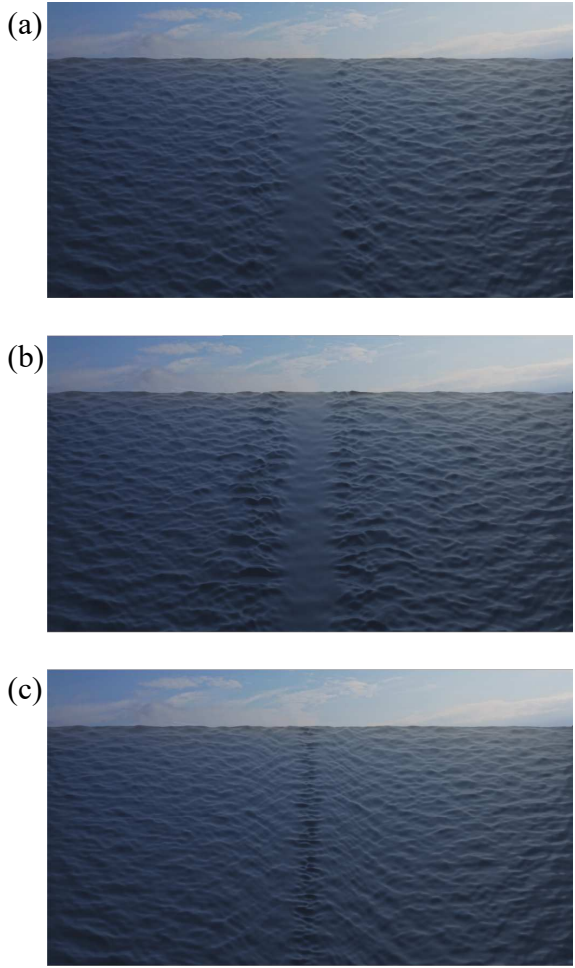}
	\caption{Top views of the ocean wave fields in the ship wake. (a) Case NR (no propeller rotating effects); (b) Case OR (propeller rotating outwards); (c) Case IR (propeller rotating inwards).  Scenes are generated using the software \textit{Blender}.} \label{fig:waves_render}
\end{figure}

\Cref{fig:waves_render} gives an overview of the ocean wave fields interacting with ocean currents for the three cases considered in this study. The modulations of waves by current variation and vortices in the wake region can be clearly observed.  In Case NR (\Cref{fig:waves_render}a), the propeller-induced vortical current is absent, and the drag-induced current velocity is the sole reason for the spatial heterogeneity of ocean-surface waves. In this case, the wave--current interaction results in a reduction in the wave steepness and a smooth wake region. \Cref{fig:waves_render}(b) shows the surface waves in Case OR, where the vortical flows rotate outwards and generate near-surface diverging flows in the wake's central region. Similar to Case NR, the region near the centerline becomes smooth. However, there exists a substantial increase in the wave steepness near the boundary between the ship-wake region and the outer region.
\Cref{fig:waves_render}(c) shows the surface waves in Case IR, in which the ship propellers rotate inwards and generate flows converging towards the center of the wake region. As a result, a narrow rough region is present at the center of the ship wake.

\begin{figure}[H]
	\includegraphics[width=0.95\columnwidth]{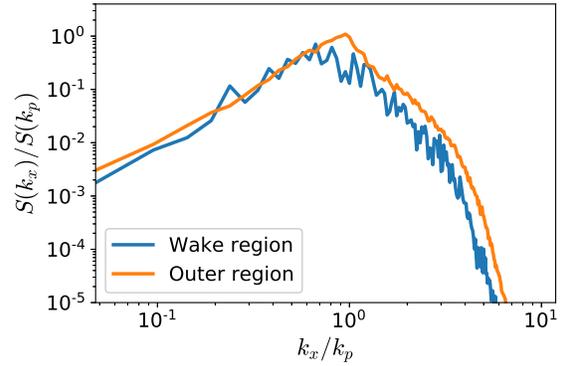}
	\caption{Wave energy spectrum $S(k_x)$ for Case NR.} \label{fig:spec_x_nov}
\end{figure}

A more detailed analysis of the change of wave steepness owing to a ship-induced current is discussed next via the one-dimensional wave energy spectrum. 
\Cref{fig:spec_x_nov} shows the one-dimensional wave energy spectrum in different regions for Case NR. Based on the profile of the drag-induced current-velocity~\cref{eq:Ud}, we define the region $L_l < y < L_r$ with a width of $L_l-L_r=20\,\mathrm{m}$ as the wake region. The region outside of the wake region is referred to as the outer region. As shown in the wake region, the one-dimensional wave-energy spectrum $S(k_x)$ is damped for almost all wave numbers. The damping effects are more significant at high wave numbers $k>k_p$. This phenomenon can be explained by the conservation of wave actions considering the current effects~\citep{longuet1961changes}. Considering a monochromatic linear wave with the original wave height $H_0$ and phase speed $C_0$ propagates into a uniform current with the speed of $U$, and the wave height $H$ becomes
\begin{align}
\frac{H}{H_0}=\frac{2}{\sqrt{1+\sqrt{1+\frac{4U}{C_0}}}\sqrt{1+\frac{4U}{C_0}+\sqrt{1+\frac{4U}{C_0}}}}.\label{eq:c_change}
\end{align}
\Cref{eq:c_change} is valid for monochromatic waves with infinitesimal wave steepness. 

For surface waves with finite amplitudes, a nonlinear correction based on wave steepness was introduced by \citet{peregrine1979finite}. In the nonlinear theory, the nondimensional averaged wave energy density $E$, averaged Lagrangian $L$, and nonlinear dispersion $S$, are rational functions of wave steepness $\epsilon=kH$, which are parameterized from the Longuet-Higgins' table of integral properties of steep waves~\citep{longuet1975integral},
\begin{align}
&E(\epsilon)=\frac{1}{2}\epsilon^2-\frac{0.19569\epsilon^4}{1-1.04488\epsilon^2-12.9792\epsilon^4},\\
&L(\epsilon)=\frac{1}{8}\epsilon^4-\frac{0.007157\epsilon^6}{1-6.73868\epsilon^2+9.64103\epsilon^4},\\
&S(\epsilon)=1+\epsilon^2+\frac{2.6107\epsilon^4(0.1935-\epsilon^2)}{1-5.63543\epsilon^2+3.98484\epsilon^4}.
\end{align}
For a wave with wave number $k$ propagating into a current with speed $U$, its wave-action density $A$ and wave-action density flux $B$ can be expressed as  $A(\epsilon)=(\rho g/\sigma k^2)(E+L)$, and $B(\epsilon)=(\rho g/2k^3)(E+5L)$. Here, $\sigma$ denotes the wave frequency relative to the current, and $\omega$ is the frequency in a fixed frame. The conservation of wave action gives
\begin{align}
A(\epsilon)U+B(\epsilon)=B(\epsilon_0),\label{eq:nonlinear1}
\end{align} 
where $\epsilon_0$ denotes the wave steepness in the absence of currents.
The nonlinear dispersion relation is 
\begin{align}
\sigma^2=gkS(\epsilon).\label{eq:nonlinear2}
\end{align}
Owing to the Doppler shift, we have the following relation,
\begin{align}
\omega=\sigma+kU.\label{eq:nonlinear3}
\end{align} 
Equations (\ref{eq:nonlinear1})--(\ref{eq:nonlinear3}) constitute a closed algebraic system for three unknowns, $\sigma$, $k$, and $\epsilon$, and the solutions can be numerically solved. Therefore, the change of wave height can be obtained. When only leading order terms with respect to $\epsilon$ are kept, the nonlinear solutions~\citep{peregrine1979finite} reduce to the linear solutions~\citep{longuet1961changes}.
When the waves propagate along with the current, i.e., $U/C_0>0$, the nonlinear-theory prediction of the wave-height change, $H/H_0$, is close to the linear-theory prediction. When the waves propagate against the current, the prediction based on the linear theory overestimates $H/H_0$. Both the Longuet-Higgins \& Stewart's solution~(\cref{eq:c_change}) and the nonlinear correction of \citet{peregrine1979finite} have been validated using numerical simulations for monochromatic waves~\citep{nwogu2009interaction,wang2018fully}. In the present study, the peak wave steepness of broadband waves is approximately $0.21$. As a result, nonlinear wave effects should be considered.

\begin{figure}[H]
	\includegraphics[width=0.95\columnwidth]{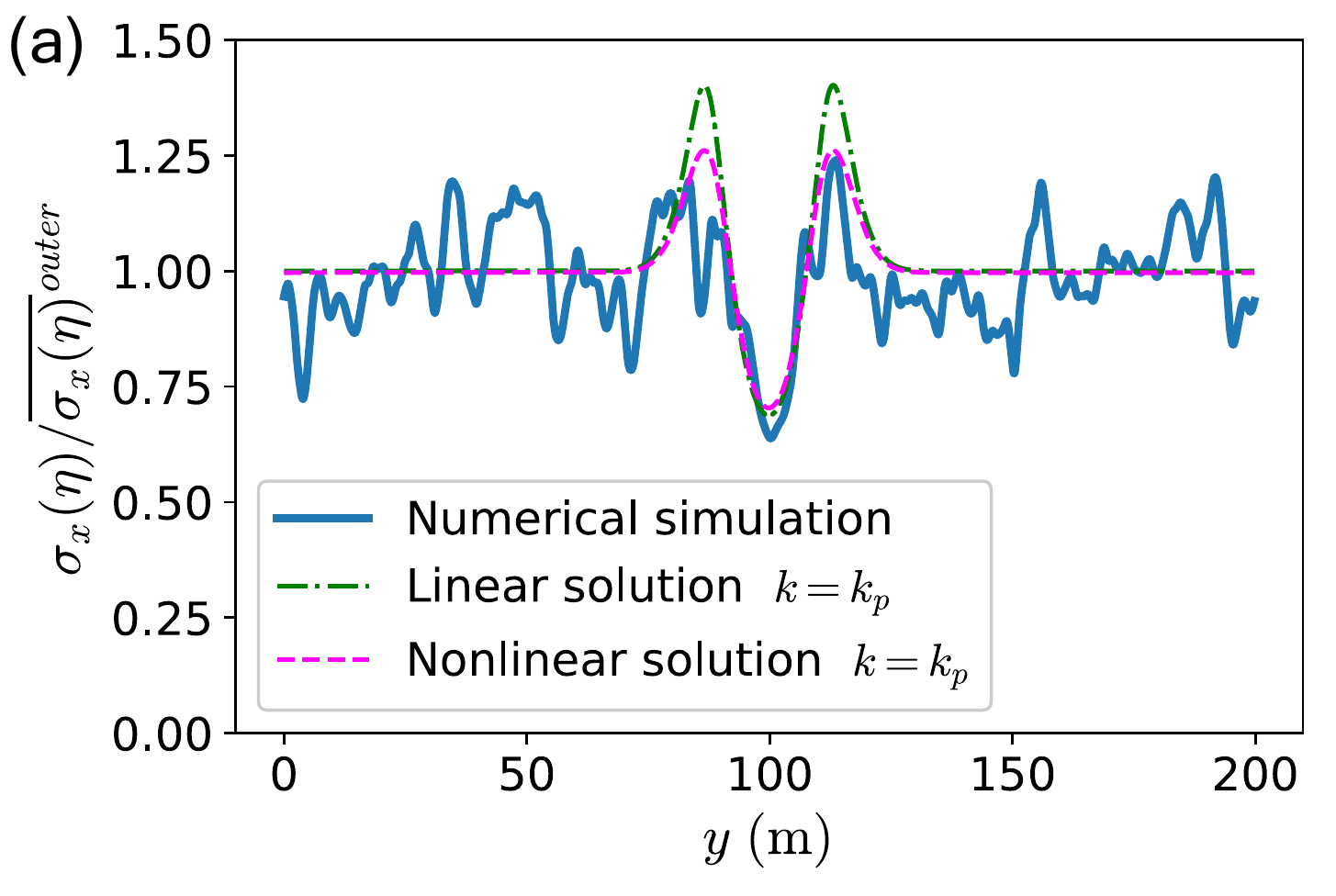}
	\includegraphics[width=0.95\columnwidth]{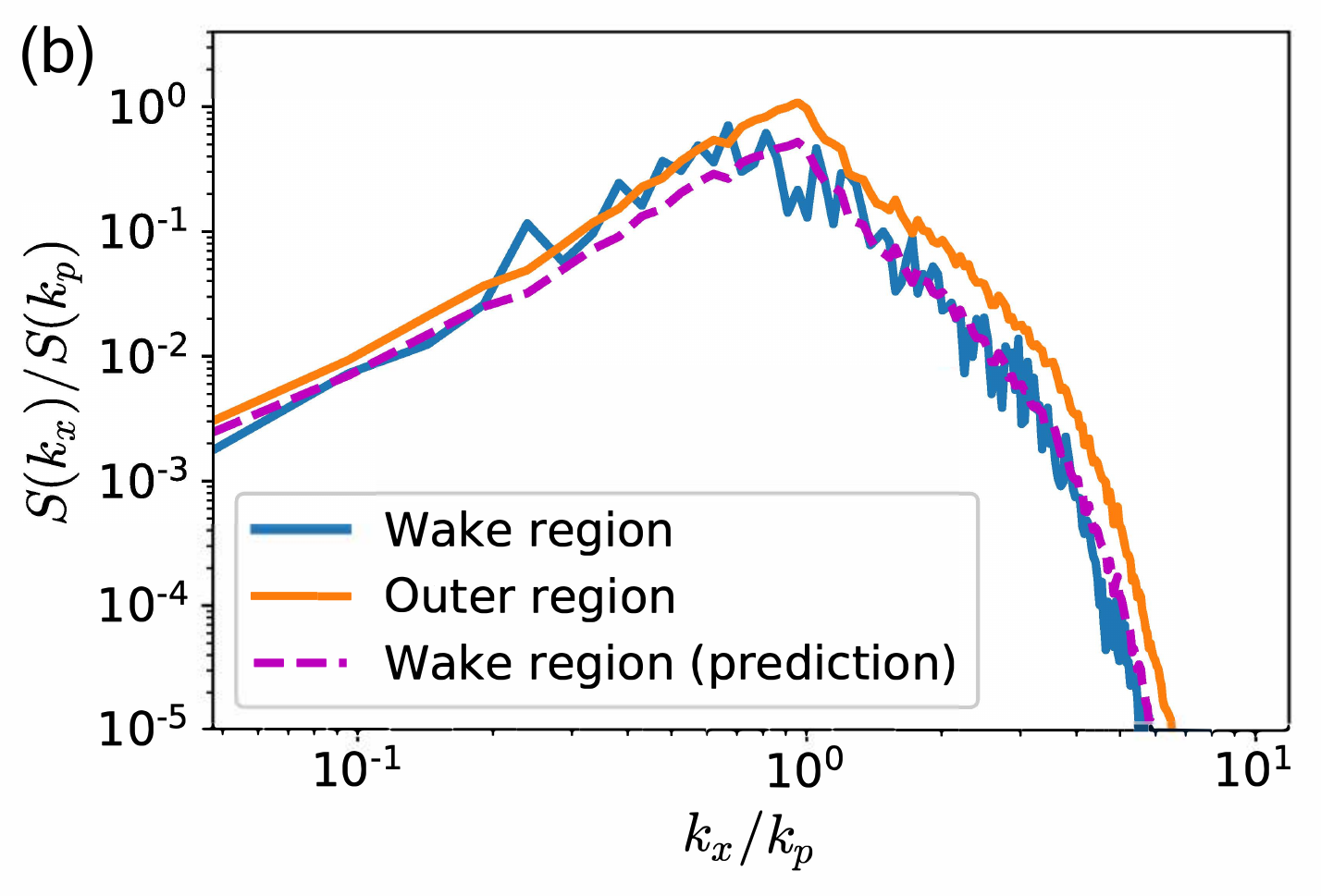}
	\caption{(a) Spanwise variation of the root mean square wave-surface fluctuations. The dash-dotted line (\textcolor{mygreen}{\textbf{--\,$\sq$\,--}}) and the dashed line (\textcolor{magenta}{\textbf{--\,--}}) represent the results based on the linear theoretical solution~\citep{longuet1961changes} and the nonlinear correction~\citep{peregrine1979finite}, respectively. (b) Comparison of the one-dimensional wave-energy spectrum between the present simulation result and theoretical prediction.} \label{fig:nov_analysis}
\end{figure}

\Cref{fig:nov_analysis}(a) shows the streamwise root mean square fluctuations of the instantaneous wave surface $\sigma_x(\eta)$ at $t=140\,T_p$ normalized using the $y$-average value in the outer region, $\overline{\sigma_x(\eta)}^{\mathrm{outer}}$. Note that $\sigma_x(\eta)$ fluctuates along the $y$-direction because the wave field is broadband and irregular. It has a global minimum at the centerline of the simulation domain $y=100\,\mathrm{m}$. The dash-dotted line (\textcolor{mygreen}{\textbf{--\,$\sq$\,--}}) is the theoretical prediction obtained by imposing the linear wave--current theory~\citep{longuet1961changes} using the peak wave number of the broadband wave field $k_p$. The linear prediction captures suppression of the wave height at the center where the waves and current are in the same direction. However, in the regions where the current is opposite to the wave-propagation direction, the amplification of the wave height is overpredicted by the linear solution. On the other hand, the solution with the nonlinear correction~\citep{peregrine1979finite}, represented by the dashed line (\textcolor{magenta}{\textbf{--\,--}}), shows a better agreement with the numerical result.

Next, we investigate the damping of the wave spectrum in the smooth region by applying the theory of \citet{longuet1961changes}~(\cref{eq:c_change}) in the wave number space. For broadband waves, we use the spectrum from the outer region and the current velocity to calculate~\cref{eq:c_change} for each wave number $k_x$ in the spectral domain and obtain the modulated wave spectrum $S(k_x)$. \Cref{fig:nov_analysis}(b) shows that the predicted spectrum (\textcolor{mymagenta}{\textbf{--\,--}}) agrees with the simulation results, especially at the high wave numbers.

\begin{figure}[H]
	\includegraphics[width=0.95\columnwidth]{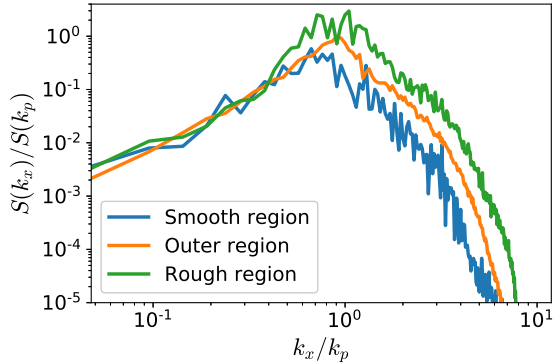}
	\caption{Wave-energy spectrum $S(k_x)$ for Case OR.} \label{fig:spec_x_div}
\end{figure}

\Cref{fig:spec_x_div} shows the one-dimensional wave energy spectrum for Case OR, in which the ship propellers rotate outwards. The entire simulation domain is empirically divided into three regions based on the observed surface roughness (\Cref{fig:waves_render}b), the smooth region $y\in[91.8\,\mathrm{m},108.2\,\mathrm{m}]$, the rough region $y\in[76.2\,\mathrm{m},91.8\,\mathrm{m}]\cup[108.2\,\mathrm{m},123.8\,\mathrm{m}]$, and the outer region $y\in[0\,\mathrm{m},76.2\,\mathrm{m}]\cup[123.8\,\mathrm{m},200\,\mathrm{m}]$.  As shown in \Cref{fig:spec_x_div}, similar to Case NR,  the energy of the high-wave number waves in the smooth region decreases compared to the outer region, indicating that the smoothness of the water surface is associated with the damping of short waves. In the rough region, $S(k_x)$ increases at both the $k\approx k_p$ and $k\gg k_p$ wave numbers. This phenomenon can be qualitatively explained by the directional-spreading property of broadband waves. For a monochromatic wave, a transverse current has little effect on wave motion. However, for broadband waves that travel in the $x$-direction, there exist wave components propagating along the $(+y)$- and $(-y)$-directions because of spreading effects (see~\cref{eq:jonswap_2d}). In this case, the diverging transverse current caused by the subsurface-vortical flow transports wave energy from the smooth region to the outer region, resulting in the energy accumulation at the boundaries of the smooth region to generate two strips where the wave steepness is much higher.

\begin{figure}[H]
	\includegraphics[width=0.95\columnwidth]{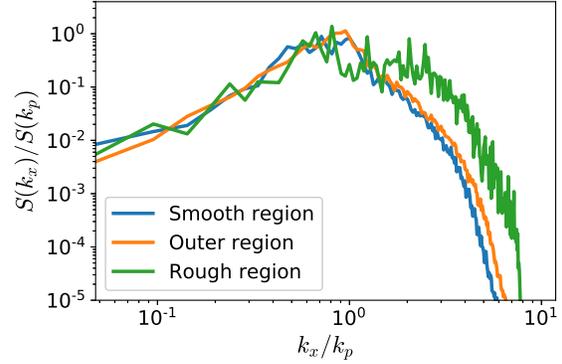}
	\caption{Wave-energy spectrum $S(k_x)$ for Case IR.} \label{fig:spec_x_conv}
\end{figure}

For Case IR, the near-surface current converges to the centerline, where wave roughness increases significantly (see~\Cref{fig:waves_render}(c)). Based on the observation, we also empirically divide the entire simulation domain into three regions; the smooth region $y\in [72.9~\mathrm{m}, 93.6~\mathrm{m}]\cup [106.1~\mathrm{m},126.8~\mathrm{m}]$; the rough region $y\in [93.6~\mathrm{m}, 106.1~\mathrm{m}]$; and the outer region $y\in[0~\mathrm{m},72.9~\mathrm{m}]\cup[126.8~\mathrm{m},200~\mathrm{m}]$. We note that the rough region, albeit narrow, has a width of $12.8~\mathrm{m}$ consisting of 64 grid points in the $y$-direction.  As shown in~\Cref{fig:spec_x_conv}, the wave-energy spectrum $S(k_x)$ in the rough region is much larger than in the other region. On the other hand, the difference between the spectra in the smooth and outer regions is small. This indicates that the inward rotating condition mainly affects waves in the narrow region along the converging line.

We use the in-house boundary-integral-based wave-current solver to numerically investigate the interactions of broadband ocean waves with ship wake flows. The numerical results capture ocean-wave characteristics under different ship-induced current conditions. On one hand, solving the evolution of ocean waves interacting with currents is a well-posed deterministic problem. On the other hand, based on the space-time observation of wave evolution, deducing the current motions underneath the free surface is an ill-posed inverse problem. In the next part, we will utilize a machine-learning algorithm to study the current-profile detection from surface-wave data.

\noindent

\section{Machine learning For Inverse Problem}
\label{SECdiscussion}
\subsection{Problem Description}\noindent
Developing an \textit{in situ} detection method for ocean currents is crucial to many applications in marine hydrodynamics and ocean engineering. Complex current velocity distributions can modulate wave dynamics in both the temporal and spatial aspects. Horizontally varying currents can generate heterogeneous surface-wave signatures owing to wave refractions. Vertically sheared currents can modify the dispersion relationship of surface waves and modulate wave behaviors in the frequency domain. Next, we study the inverse problem.

In the present inverse problem, the main goal is to deduce the current field based on the observation of surface wave data, with the nonlinearity of the wave--current interaction accounted for.  Recently, deep learning has shown great potential in many areas, such as image recognition and turbulence modeling~\citep{lecun2015deep,duraisamy2019turbulence}.  In this work, we have developed a data-driven deep-learning framework to solve the inverse problem of current detection based on surface-wave data.  The framework takes wave data as the input and outputs the information of the current underneath, including its magnitude and spatial distribution.  To evaluate the performance of the deep learning algorithm, we consider the canonical problem of a wave group interacting with a shear current.  
The definition of this inverse problem is introduced using a statistical learning framework~\citep{shalev2014understanding} as follows.
The \textit{Domain Set} $\mathcal X$, defined as a set of objects that we would like to regress, is the wave elevation, which is discrete in both space and time. The \textit{Target Set} $\mathcal Y$ is a set of target outputs, which is a set of parameters of ocean currents in our settings. We can assume that there exists a distribution $D$, which characterizes the pairs $z=(x,y)\in\mathcal{X}\times\mathcal{Y}$.  The \textit{Hypothesis} is a class of functions $\mathcal H=\{h~|~h:\mathcal X\rightarrow\mathcal Y\}$ that provide the predictions. The main goal of the machine-learning method is to find an optimal function in the hypothesis class to describe the distribution $D$. To measure the distance between the predicted result $h(x)$ and the target $y$, we need to define the loss function $l(h,z)$. In the present regression problem, the square loss function is adopted,
\begin{align}
l(h,z)=l(h,(x,y))=(h(x)-y)^2.
\end{align}
The goal of the training algorithm in the machine learning process is to use the \textit{Training Data} $S=\{z_i=(x_i,y_i)\in\mathcal{X}\times\mathcal{Y}\,|\,i=1,\dots,m\}$, a set of finite pairs with size $m$, to minimize the following training error $L_S(h)$,
\begin{align}
L_S(h)=\frac{1}{m}\sum_{i=1}^m l(h,z_i).
\end{align}
Clearly, we cannot access any information of the distribution $D$, and we hope that the algorithm learned on the training data $S$ can be generalized to reveal the properties of the distribution $D$. The corresponding generalization error function $L_D(h)$ over the distribution $D$ is defined as
\begin{align}
L_D(h)=\mathrm{E}_{z\sim D}[l(h,z)].
\end{align}
In the present study, we aim to explore the capability of the machine-learning method to inversely calculate ocean-current distribution based on surface-wave data.
In a realistic ocean environment, the wave steepness can be sufficiently large such that the nonlinear effect should be taken into consideration. Moreover, the ocean current might be heterogeneous in three dimensions and might vary in time, which further complicates the inverse deduction. Here, as a first step of the study, we simplify the problem and only consider linear waves propagating on a linearly sheared uniform current. The advantage of this simplification is that we can generate arbitrarily large data samples based on the theoretical solution. For more complex current conditions and finite-amplitude wave effects, we can use data from numerical simulations or experimental measurements to develop the machine-learning algorithm, although the cost to obtain such datasets is much higher. Because the machine-learning technique is a data-driven method, it is feasible for the model discussed in this study to be generalized to more realistic conditions.

\subsection{Generation of Datasets}
We will now consider small-amplitude, one-dimensional water waves propagating on a two-dimensional linearly sheared current. Let $(\xi,\zeta)$ denote the two-dimensional Cartesian coordinates where $\zeta=0$ is located at the mean water surface. The surface elevation of the one-dimensional wave group is given by
\begin{align}
\eta(\xi,\tau)=\sum_{i=1}^N a_i\sin(k_i \xi-\omega_i \tau+\phi_i).\label{eq:def_wava_group}
\end{align}
The velocity of the deep-water sheared current is expressed as
\begin{align}
U(\zeta)=U_0+\zeta d, \quad \zeta\in(-\infty,0].
\end{align}
Here, because the wave amplitude is assumed to be small, the wave group can be considered as a linear combination of Airy waves with a different wave number $k$, i.e., the nonlinear wave--wave interaction and the energy transfer among different wave numbers can be neglected. In~\cref{eq:def_wava_group}, for each wavenumber $k_i$, the variables $a_i$, $\omega_i$, $\phi_i$ represent the corresponding wave amplitude, angular frequency, and phase, respectively. The dispersion relation of the water waves is highly nonlinear with respect to the current-velocity distribution. When the current velocity is linearly sheared vertically, the wave angular frequency $\omega_i$ can be determined from the following dispersion relation,
\begin{align}
\omega_i=U_0 k_i-\frac{d}{2}+\sqrt{\left(\frac{d^2}{4}+gk_i\right)},
\end{align}
where $g$ denotes the gravitational acceleration. In this model, we set $g=10~\mathrm{m/s^2}$ for simplicity,

From measurements and simulations, we can only create or collect discrete data. Therefore, in the machine-learning approach, it is natural to choose a discrete-in-time and discrete-in-space dataset from the wave--current interaction problem for the training algorithm to learn. The datasets generated in this study use a spatial domain of length $2\pi$ discretized uniformly by 128 grid points. The total number of wave modes is set to be $N=20$, i.e., we have the wave number $k_i=i,\,i=1\dots N$. Note that the wave amplitude $a_i$ is associated with the wave energy. To mimic the decay of wave energy in high frequencies in nature, we define the following weighted norm,
\begin{align}
\left\Vert\eta\right\Vert_W=\sum_{i=1}^Nf(i)a_i^2,
\end{align}
with the (arbitrarily designed) weighted function
\begin{align}
f(\gamma)=10+0.2\gamma^2.
\end{align}
The generation of $a_i$ for different samples is conducted by drawing from a uniform distribution with the normalization restriction $\left\Vert\eta\right\Vert_W=1$. In this model, we restrict the range of surface current velocity $U_0$ and the shear rate $d$ to
\begin{align}
U_0\in[0,1],\\
d\in[0,1].
\end{align}
To generate the target set $y_i=(U_{0i},d_i)$, we assume that $\{y_i\}$ satisfies the uniform distribution $\mathcal{U}[0,1]^2$. The phase $\phi_i$ is also assumed to follow the uniform distribution $\mathcal{U}[0,2\pi]$.
For each wave component with the wave number $k$, the ratio between the surface-current velocity and phase velocity, $U/C$, varies from $0$ to $M(k)$, where $M(k)\in[0.32,0.64]$ for $k\in[1,20]$. As for the temporal discretization, we set the observation interval $\delta=0.2$ and collect 10 consecutive segments $\eta|_{t=t_j}$ for $j=0,\dots 9$. The temporal-spatial input data are then rearranged into one-dimensional arrays.
The above process can generate independent and identically distributed (IID) random variables $\{\mathcal{S}=\{z_i=(x_i,y_i)\}$, which are used as the training data. In this study, the IID training data consists of $m=5\times 10^5$ different samples. For each sample of training data $\bm z_i=(\bm x_i,\bm y_i)$, the dimensions of $x_i$ and $y_i$ are $1024\times1$ and $2\times1$, respectively. To describe the unknown distribution $D$, we generate another IID dataset with a size of $m=10^6$ as the test dataset to approximate the distribution.

Both the training and test dataset represent the evolution of broadband waves with 20 adjacent wave modes over linearly-sheared currents. The surface current velocities $U_0$ and the shear rates $d$ for all cases recover a continuous parameter space $U_0\times d\in [0,1]^2$ due to the central-limit theorem.

\subsection{Network Structure and Training Algorithm}
We use multi-layer, fully-connected neural networks to map the input $x_i$ of the size $1280\times1$ to the output $y_i$ of the size $2\times1$. The prediction function $h(\bm x)$ of the multi-layer network is defined by the following recursive representation,
\begin{align}
&h(\bm{x})=\bm {\mathrm{a}}\bm{x}^{(H)}+\bm{b},\\
&\bm{x}^{j}=\sigma\left(\bm{\mathrm{W}}^{(j)}\bm{x}^{(j-1)}+\bm{b}^{(j)}\right),\,1\leq j<H
\end{align}
where $H$ is the number of hidden layers; $\bm {\mathrm{W}}^{(j)}\in\mathbb R^{d_{j-1},d_j}$ and ${\bm b}^{(j)}$ are the weight and bias at the $j$-th hidden layer, respectively; $\bm {\mathrm{a}}$ and $\bm b$ are respectively the weight and bias of the output layer; and $\sigma$ is the element-wise activation function. In this study, we choose the rectified linear unit (ReLU) function $\sigma(x)=\max\{0,x\}$ as the nonlinear activation function~\citep{glorot2011deep}. \Cref{Tab:mlp} summarizes the information about the number of hidden units $d_j$ in each hidden layer $j$ of the neural network. The total number of elements in this neural network is approximately two million.

\begin{table}[H]
	\caption{Details of the neural network}
	\begin{small}
		\begin{center}
			\begin{tabular}{|c|c|c|c|c|c|c|}
				\hline
				\makecell{Hidden layer\\$j$} & $1$ &$2$ & $3$ & $4\sim 8$ & $9$ &$10$ \\
				\hline
				\makecell{Hidden units\\ $d_j$} & $800$ & $600$&$400$& $300$&$100$&$64$\\
				\hline
			\end{tabular}
		\end{center}
	\end{small}
	\label{Tab:mlp}
\end{table} 

We use the back-propagation method to find the minimizer $h$ of the loss function $L_S(h)$. The initialization of the weight and bias matrices is an essential point to train the network. In this study, we use the normal distribution $\mathcal{N}(\mu,\sigma)$ with the mean $\mu=0$ and the standard deviation $\sigma=0.03$ to randomly initialize each weight and bias. The mini-batch gradient-descent method~\citep{bertsekas1996incremental} is adopted for determining the gradients in the back-propagation. This optimization method can accelerate the training process significantly compared to the stochastic gradient-descent method. Moreover, by choosing the appropriate batch size, this method can handle arbitrarily large training data, overcoming the limit of the device memory. For our training process, the batch size is chosen as $2\times10^4$, and the learning rate of the gradient-descent method is $\eta=0.035$ at the beginning and decreases to $\eta=0.001$ when the training epochs reach $60$. The total training epochs are set to be $600$. The mini-batch gradient-descent method is performed using the package \textit{TensorFlow} deployed on an NVIDIA TESLA V100 GPU.

\begin{figure}[H]
	\includegraphics[width=0.95\columnwidth]{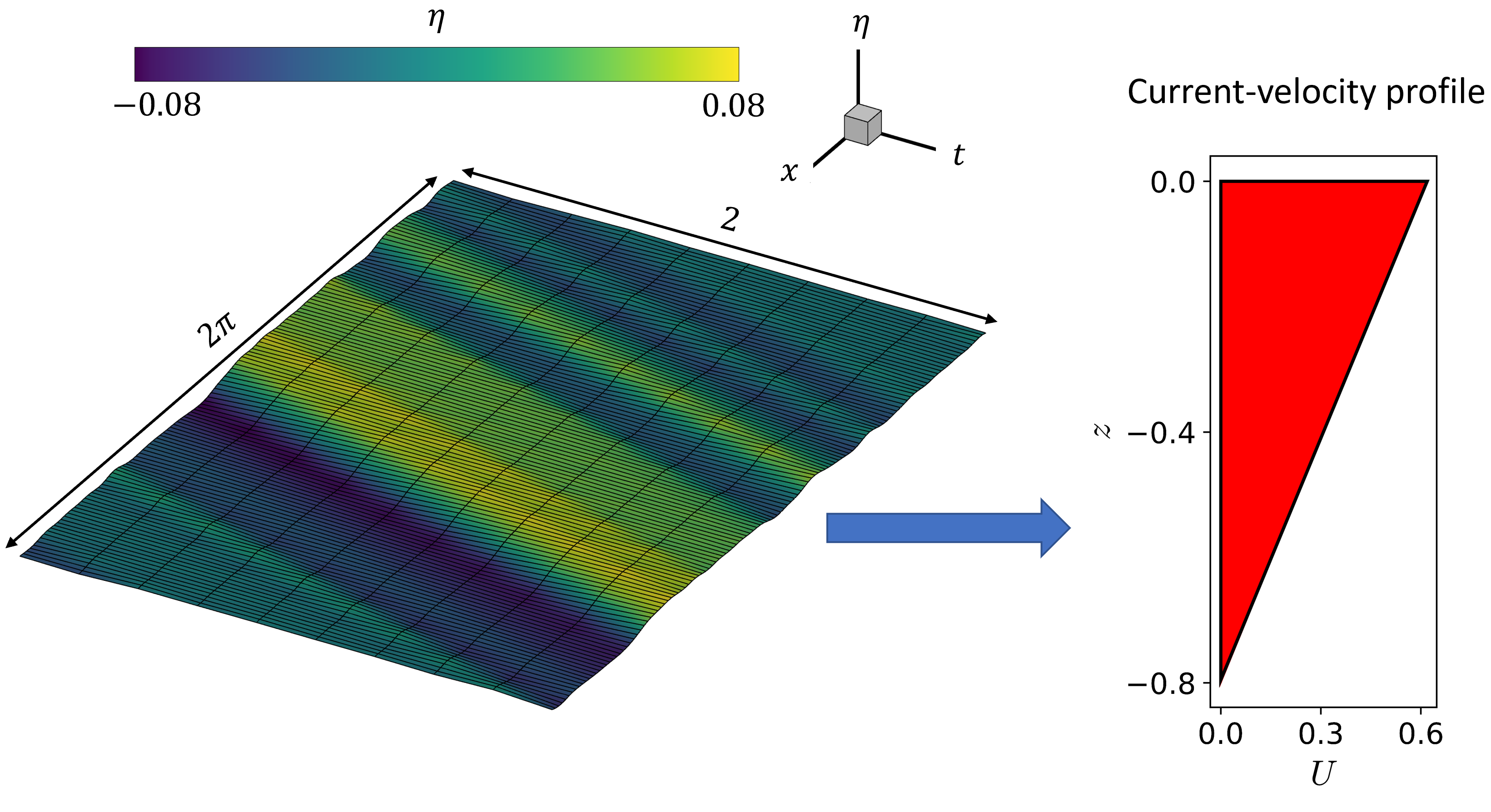}
	\caption{Illustration of the proposed machine-learning input-output data structure. The left figure shows one example of the input space-time surface wave data from the test set. The right figure shows the current-velocity profile. The red area represents the ground truth of current-velocity distribution, and the black edges represent the predicted results from the machine-learning algorithm.} \label{fig:illustration}
\end{figure}

Figure~\ref{fig:illustration} shows an example of the input data and output data from the test dataset in the machine-learning algorithm. The left contour is a visualization of a space-time surface-wave data from the test dataset. The right figure shows the current-velocity profile. The red area represents the ground truth of current-velocity $U(z)$, and the black edges denote the predicted current-velocity profile using the machine-learning algorithm. In this example, once a neural network is trained, the relative errors between the predictions and ground truth for surface velocity $U_0$ and shear rate $d$ are only $0.43\%$ and $0.88\%$, respectively.
\begin{figure}[H]
	\includegraphics[width=0.95\columnwidth]{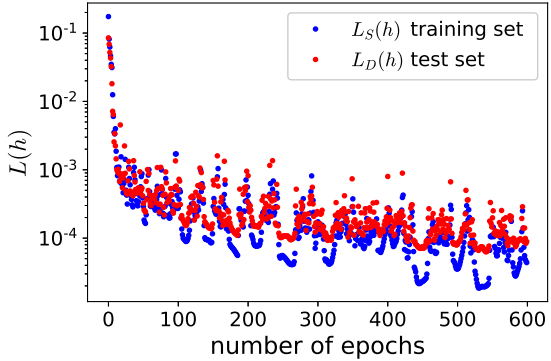}
	\caption{Training process of the algorithm.} \label{fig:training_once}
\end{figure}

\subsection{Results}
\Cref{fig:training_once} shows an example of the training process. The blue dots represent the training error $L_S(h)$ based on the training data. The red dots represent the generalization error function $L_D(h)$, which can be approximated by the test dataset with $10^6$ samples according to the central-limit theorem. In \Cref{fig:training_once}, both the training error $L_S(h)$ and the generalization error $L_D(h)$ follow an oscillatory and exponential decay as the number of epochs increases. When the epoch reaches 600 and we stop the training process, the final training error and generalization error reach $L_S(h)=4.4\times10^{-5}$ and $L_D(h)=8.9\times10^{-5}$, respectively. No error saturation has been observed during the training process. These results show the capability of the designed machine learning algorithm to inversely calculate the ocean current distribution based on the surface-wave profiles.

Realistically, iterations of back-propagation can only be run for a finite number of steps because of computational cost. We expect the non-zero residual errors for both the training set and the test set and the distributions of residual errors to be closely related to the performance of the deep neural network model. Therefore, the uncertainty quantification for the algorithm should be carefully analyzed. The mean-square error definition of the loss function $L(h)$ is widely used in regression problems. However, the mean-square error cannot explicitly control the pointwise error of the dataset. We are curious about how the error function $h(x)-y$ is distributed in the set $\mathcal{X}\times\mathcal{Y}$ when the loss function $L(h)$ is well controlled within a small threshold by the training algorithm. The events when $h(x)-y$ is far away from its mean value are of more interest. If we have $m$ samples in the dataset, the following norm inequality holds,
\begin{align}
\Vert  h(x)-y \Vert_\infty \leq \sqrt{m} \sqrt{L(h)}.
\end{align}
It means that the maximum absolute value of $h(x)-y$ is bounded by the loss function $L(h)$ and the total number of samples $m$. Even though the loss function $L(h)$ can be decreased by increasing the number of samples of the training set $m$ and the total number of training epochs, as $m$ increases, the upper bound of $\Vert  h(x)-y \Vert_\infty$ also grows. For any data-driven model, increasing the data samples is always helpful to obtain good results.
However, due to the increase of the upper bound, it is reasonable to expect that the rare event when $h(x)-y$ is far away from its mean value may occur.

We compute the pointwise error $r(h_s,z)=h(x)-y$ at each $(x,y)$ pair in both the training set $z\sim S$ and the test set $z\sim D$. For any $\beta>0$, we can define two probabilities for the rare events as,
\begin{align}
&\mathrm{Pr}_{z\sim S}\left[\left|\left(|r(h_s,z)|-\mu_S\right)\sigma_S^{-1}\right|>\beta\right]\notag\\
&=\frac{1}{m_S}\sum_{z_i\sim S} \bm 1_{|r(h_s,z_i)|\sigma_S^{-1}>\beta},\\
&\mathrm{Pr}_{z\sim D}\left[\left|\left(|r(h_s,z)|-\mu_D\right)\sigma_D^{-1}\right|>\beta\right]\\
&=\frac{1}{m_D}\sum_{z_i\sim D} \bm 1_{|r(h_s,z_i)|\sigma_S^{-1}>\beta},\notag
\end{align}
where $\sigma_S$ and $\sigma_D$ denote the standard deviations of $|r(h_s,z_i)|$ in the training set $S$ and test set $D$, respectively, $m_S$ and $m_D$ denote the total number of samples in the training set and test set, respectively, and $\bm 1_{x>\beta}$ is the indicator function.

\begin{figure}[H]
	\includegraphics[width=0.95\columnwidth]{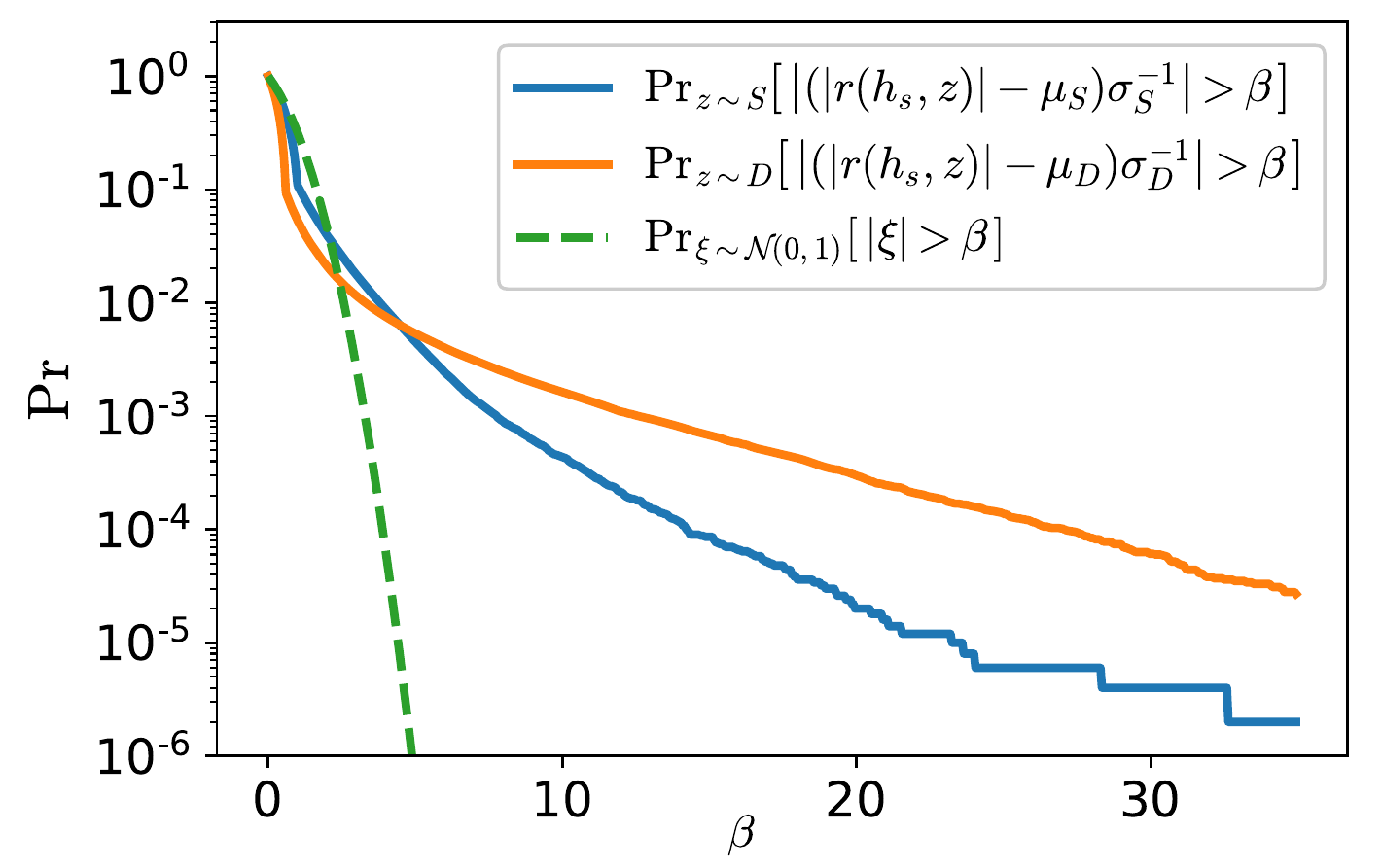}
	\caption{Probability of rare events.} \label{fig:long_tail}
\end{figure}

\Cref{fig:long_tail} shows the probabilities when $|r(h_s,z_i)|$ lies outside of the range $[\mu-\beta\sigma,\mu+\beta\sigma]$ against different $\beta$ for both the training set (blue line \textcolor{myblue}{\textbf{-----}}) and the test set (orange line \textcolor{myorange}{\textbf{-----}}). The green dashed line shows the results of a normal distribution, where $\mathrm{Pr}=1-\mathrm{erf}(\beta/\sqrt{2})$.
The tails of the distributions for both the training set $S$ and the test set $D$ are exponentially bounded and are much heavier than the tail of a normal distribution. For a normal distribution, the probability $\mathrm{Pr}$ of an event that its normal deviate lies out of the range $\mu -\beta\sigma \leq r \leq \mu +\beta\sigma$ is below $10^{-6}$ when $\beta=5$. However, such probabilities with $\beta>5$ for the training set and test set are much larger, indicating that an extremely rare event is more likely to occur in the current deduction problem. For the training set of the size $5\times10^5$, we observe that there is one sample satisfying $||r(h_s,z)|-\mu_S|>35\sigma_S$, which gives the probability $\mathrm{Pr}=2\times10^{-6}$. This extremely rare event is expected to occur because the loss function for the regression is the mean square error, which only evaluates the error in an average sense over the entire data. We also note that the distribution tail of the test set is heavier than the tail of the training set. This result is reasonable because we do not access any samples in the test set during the training process. The tail for the test set decays as $\exp(-\beta)$, which still supports the generalization of the machine learning algorithm to the entire distribution $D$. Inevitably, there is still a relatively low possibility that some samples may have large generalization errors.

\begin{figure}[H]
	\includegraphics[width=0.95\columnwidth]{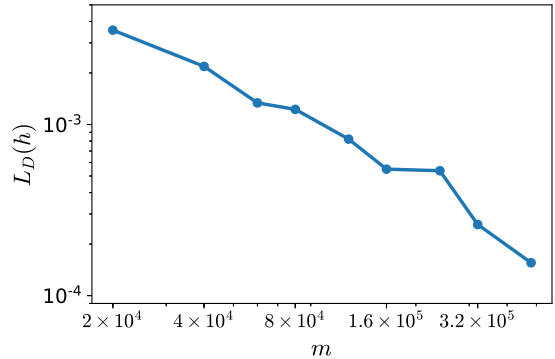}
	\caption{Dependency of the generalization error $L_D(h)$ on the size of training set $m$.} \label{fig:test_set_m}
\end{figure}

In the above discussions, we have shown that a sufficiently extensive training dataset can allow extreme events to occur, but this does not mean that increasing the size of training data is an improper option. Next, we investigate the effect of size $m$ of the training set on the generalization error $L_D(h)$, and show that increasing the size of training data can decrease the generalization error. We use the same test set of the size $10^6$ to represent the true distribution $D$ and choose nine different training sets of sizes $m$ ranging from $2\times10^{4}$ to $4.8\times10^{5}$.
For the small training set case ($m=2\times10^4$), overfitting the training set is observed because the training error continues decreasing while the test error is stabilized. Even though in the late training process there are large gaps between the training error and test error, overfitting does not affect the generalization of the algorithm~\citep{kalimeris2019sgd}. 
\Cref{fig:test_set_m} shows the dependency of the generalization error $L_D(h)$ on the size of training set $m$. The initialization of weights and biases satisfies the same normal distribution $\mathcal{N}(0,0.03)$ for different cases. All the training sets are trained by the back-propagation method with a total of $600$ epochs. Then we evaluate the generalization error of each training set on the same large test set with $m=10^6$. Both axes in \Cref{fig:test_set_m} are in the logarithmic scale, and a power-law relationship  $L_D(h)\sim m^{-1.1}$ is found. This phenomenon supports the argument that increasing the training set size is helpful to obtain a smaller generalization error, yet the trade-off between the expected generalization error and computational cost still needs to be considered. 
From this study, the tendency of the effect of the training size on the generalization error provides a guidance for choosing a suitable training-set size in engineering applications.

In the present study, we consider the linear dispersive properties of water waves and constraint the function space of current velocity. These assumptions can benefit us to fast generate a large dataset with IID samples. The total number of grid points in the test dataset is over 1 billion. 
The idea of artificial neural networks has been proposed over several decades. However, with the boost of computing powers, deeper and wider networks are revealing a promising capability to deal with complex tasks. For example, the Generative Pre-trained Transformer 3 (GPT-3), a 175 billion parameter autoregressive language model, was trained using a dataset with roughly 400 billion byte-pair-encoded tokens and GPT-3 shows that scaling up the sizes of network models and datasets significantly improves the task performance~\citep{brown2020language}. Our present study evaluates the pointwise prediction error of a trained deep neural network in a statistical aspect and investigates the effect of training-data size on the task performance in a large test dataset. We address the necessity of statistical accuracy evaluation of the trained network and the trade-offs between the precision of network models and the computing cost in the training process.
Our further study will focus on the generalization of present machine-learning approach to study interactions between two-dimensional broadband waves and arbitrarily-distributed current field and deduce the underneath current motions based on observations of surface waves.

\section{Conclusions}
\label{SECconclusion}

In this study, we have performed direct phase-resolved wave simulations under complex ocean-current conditions, and numerically investigated the effects of horizontally-sheared current and vortical flow on broadband surface waves. The simulations capture the spatial variations of ocean-wave roughness caused by the influence of a spatially-varying current field. The local changes of wave steepness owing to wave--current interactions are quantitatively explained with existing theories.
We have also developed a deep-learning framework to inversely deduce the current field based on the temporal-spatial discrete wave data. 
Without any \textit{a priori} physics-based knowledge, the machine-learning algorithm can recover the interactions between one-dimensional broadband waves and vertically-sheared current.
The capabilities of a machine-learning algorithm on minimizing errors in the training dataset and its generalization error on the test dataset are examined. The distribution of the deviation of the predictions from the actual ocean-current profile over the entire dataset is analyzed. It is found that the distribution of the deviation is not normal and its tail decays only exponentially. The power-law dependency of the generalization error on the training-data size is observed via numerical experiments.
The results presented in this work show that high-fidelity simulation methods and the deep-learning-based inverse modeling approach are promising for applications in naval hydrodynamics research.

\section{Acknowledgements}
\label{SECacknowledgements}

Some of the results were obtained from the support by the Advanced Naval Platforms Division of the Office of Naval Research through the project “Simulation Interface between Marine Environment and Sea Platforms” (N00014-19-1-2139) managed by Dr. Peter Chang.

\bibliographystyle{SNHstyle}
\bibliography{33SNH}
\end{multicols*}
\end{document}